\documentclass[sigconf]{acmart}
\usepackage{dsfont}
\usepackage{adjustbox}
\usepackage[utf8]{inputenc}
\usepackage{amsmath,amssymb}  
\usepackage[mathscr]{eucal}
\AtBeginDocument{%
  \providecommand\BibTeX{{%
    \normalfont B\kern-0.5em{\scshape i\kern-0.25em b}\kern-0.8em\TeX}}}

\setcopyright{acmcopyright}
\copyrightyear{2020}
\acmYear{2020}
\acmDOI{10.1145/3366424.3386198}

\copyrightyear{2020}
\acmYear{2020}
\acmConference[WWW '20 Companion]{Companion Proceedings of the Web Conference 2020}{April 20--24, 2020}{Taipei, Taiwan}
\acmBooktitle{Companion Proceedings of the Web Conference 2020 (WWW '20 Companion), April 20--24, 2020, Taipei, Taiwan}
\acmPrice{}
\acmDOI{10.1145/3366424.3386198}
\acmISBN{978-1-4503-7024-0/20/04}

\begin{document}

\title[“An Image is Worth a Thousand Features”]{“An Image is Worth a Thousand Features”: Scalable Product Representations for In-Session Type-Ahead Personalization}

\author{Bingqing Yu}
\authornote{Authors are ordered by direct contribution - idea and execution - in the research project.}
\affiliation{%
  \institution{Coveo}
  \city{Montreal}
  \country{Canada}}
\email{cyu2@coveo.com}

\author{Jacopo Tagliabue}
\authornotemark[1]
\authornote{Corresponding author.}
\affiliation{%
  \institution{Coveo Labs}
  \city{New York}
  \state{NY}
}
\email{jtagliabue@coveo.com}

\author{Ciro Greco}
\authornotemark[1]
\affiliation{%
    \institution{Coveo Labs}
    \city{New York}
  \state{NY}}
\email{cgreco@coveo.com}

\author{Federico Bianchi}
\authornotemark[1]
\affiliation{%
  \institution{Bocconi University}
  \streetaddress{Via Roberto Sarfatti, 25 }
  \city{Milano }
  \country{Italy}}
\email{f.bianchi@unibocconi.it}

\renewcommand{\shortauthors}{Yu et al.}

\begin{abstract}
  We address the problem of personalizing query completion in a digital commerce setting, in which the bounce rate is typically high and recurring users are rare. We focus on in-session personalization and improve a standard noisy channel model by injecting dense vectors computed from product images at query time. We argue that image-based personalization displays several advantages over alternative proposals (from data availability to business scalability), and provide quantitative evidence and qualitative support on the effectiveness of the proposed methods. Finally, we show how a shared vector space between similar shops can be used to improve the experience of users browsing \textit{across} sites, opening up the possibility of applying zero-shot unsupervised personalization to increase conversions. This will prove to be particularly relevant to retail groups that manage multiple brands and/or websites and to multi-tenant SaaS providers that serve multiple clients in the same space.
\end{abstract}

\ccsdesc[500]{Information systems~Query suggestion}
\ccsdesc[300]{Information systems~Language models}
\ccsdesc[300]{Information systems~Personalization}

\keywords{e-commerce query auto-completion, conditional language model, neural networks, zero-shot learning, transfer learning}

\maketitle

\section{Introduction}
Type-ahead systems have a long distinguished history in Information Retrieval \cite{Cai:2016:SQA:3099948} and are a crucial component of all contemporary state-of-the-art commerce platforms: from the point of view of the retailer, suggesting queries to shoppers when they type in the search bar has the benefit of nudging them into querying for \textit{what} the shop wants them to buy \textit{in the way} the shop wants.

The recent wave of adoption of neural architectures in commerce (\cite{Grbovic15}, \cite{Vasile16}) has changed type-ahead systems as well (\cite{inproceedingsparj2017}); \textit{this} work presents a simple-to-implement, principled and effective way to inject real-time personalization into query suggestions in two distinct commerce scenarios: first, we tackle the \textit{within shop} personalization challenge, showing how dense features calculated from product images allow us to quickly compute session vectors that successfully capture shopping intentions; second, we tackle the \textit{cross-shop} personalization challenge, showing how the same deep learning architecture can leverage a shared vector space of product images to transfer the intent from website \textbf{X} to type-ahead on website \textbf{Y}, even when \textbf{X} and \textbf{Y} only have partially overlapping items (as a more extreme case, even when \textbf{X} and \textbf{Y} are in completely different languages). 

The \textit{cross-shop} scenario is a typical case of ``transfer learning'', and it is a particularly important one for multi-brand groups and, generally, A.I. providers scaling predictions across their network as representing user preferences in type-ahead systems is a formidable challenge for typical mid-size digital shops. The combination of high bounce-rate, low ratio of recurring users, small volume of search queries and inconsistent meta-data make in-session personalization both hard to achieve and \textit{extremely valuable}, as it promises to improve the user experience of a large pool of shoppers and turn them into buyers. Using industry data, we argue that \textit{within shop} and \textit{cross-shop} personalization is a crucial problem to solve in today's competitive market of digital retailers, and provide substantial evidence that the proposed methods are a substantial improvement over industry best practices. We summarize the main contributions of \textit{this} paper in two major points:
\begin{itemize}
\item we extend the noisy channel model for type-ahead, by "injecting" personalization into the language model using dense vectors from product images; we present methods that are easily applicable even to websites with no historical fine-grained user tracking, or languages for which no linguistic resources exist;
\item we tackle the problem of scaling personalization \textit{across websites}, as for example across two brands in the same group, or two versions of the same site with different localization: in particular, we show how shoppers' interest can be successfully transferred between websites \textbf{X} and \textbf{Y} (with no prior data point for the user on \textbf{Y}) by leveraging the shared vector space.
\end{itemize}
\textit{This} work is structured as follows: Section \ref{use_section} details our two main motivating use cases from the e-commerce industry; Section \ref{related_section} discusses relevant previous work on type-ahead, with special focus on neural models; Section \ref{dataset_section} gives a high-level description of the target dataset; Section \ref{method_section} first introduces two ways in which we can evolve the noisy channel model to deal with contextual elements and then explains how to calculate a context vector. Finally, we benchmark the models in Section \ref{experiment_section} and provide  quantitative and qualitative assessment of the quality of the predictions, before concluding in Session \ref{conclusion_section} with our product roadmap to engineer these models at scale.

\section{TYPE-AHEAD IN PRODUCT SEARCH: A MOTIVATING USE CASE}
\label{use_section}
\begin{figure}[h]
  \centering
  \includegraphics[width=\linewidth]{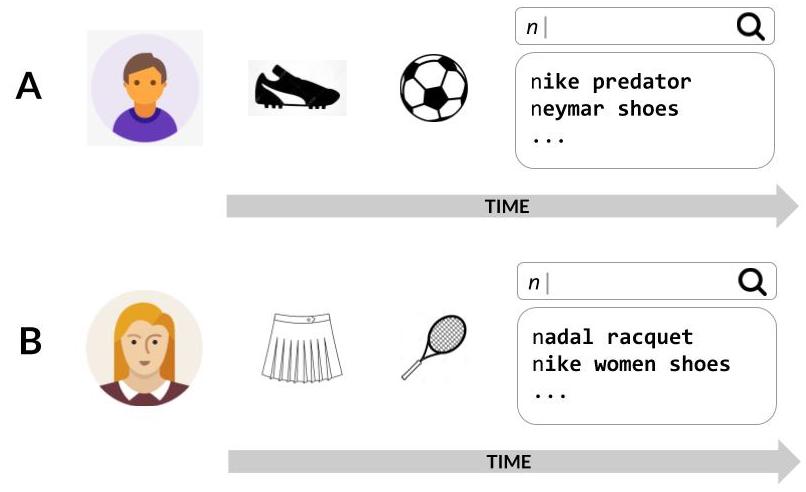}
  \caption{Shopper \textbf{A} and shopper \textbf{B} browse a sport apparel website and then start typing in the search bar: their shopping intent should somehow influence the ranking of possible query completions. }
  \label{use_pic}
  \Description{Type-ahead services need personalized ranking to be relevant.}
\end{figure}

Consider the browsing patterns of \textbf{Shoppers A} and \textbf{B} in Figure \ref{use_pic}, visiting a digital shop selling sport apparel: \textbf{Shopper A}'s latent intent is about “soccer”, while \textbf{Shopper B}'s is about “tennis”. When they type the character “n” in the search bar, the type-ahead service needs to provide them with relevant suggestions, “soccer”-based for \textbf{A} and “tennis”-based for \textbf{B}. In other words, the ideal language model behind the scene is a \textit{conditional} language model: $P("nadal" | "n")$ is $P("nadal" | "n", I)$, where \textit{I} is the latent intent of the current user. To strengthen this point, we report in Table \ref{tab:example_queries} some examples of real query from our dataset (Section \ref{dataset_section}): the amount of overlap in the top queries across website sections is negligible, making "global" probability estimates fairly inaccurate for many shoppers. 

\begin{figure}[h]
  \centering
  \includegraphics[width=\linewidth]{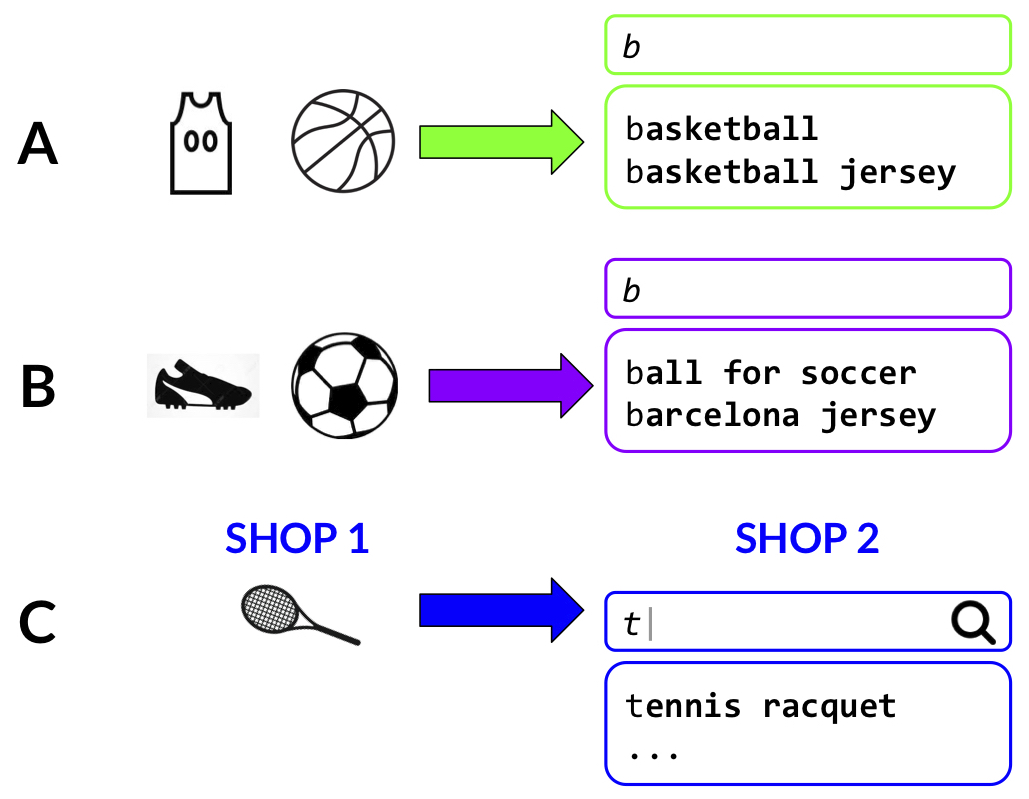}
  \caption{Shopper \textbf{C} is browsing soccer-related products on \textbf{Site 1}, then starts a browsing session on a related site (\textbf{Site 2}). Can we provide personalized type-ahead completions starting from the very first interaction?}
  \label{pred_pic}
  \Description{Zero-shot type-ahead personalization.}
\end{figure}

Personalization within a session is certainly important, but there is another, often neglected type of personalization - what we call “zero-shot personalization”. Consider now \textbf{Shopper C} in Figure \ref{pred_pic}: \textbf{C} starts browsing soccer products in \textbf{Site 1} and then move to \textbf{Site 2} - a different e-commerce site selling sport apparel. \textbf{C} has never been on \textbf{Site 2}, but we would still like to power her experience with something more than a generic language model for query suggestions: however, we do not have data points for her on \textbf{Site 2} and the (unconditioned) language model in \textbf{2} may differ significantly from the one on \textbf{Site 1}. In an industry where many traditional companies struggle to keep up with the personalization strategies of the top tech players in the space, finding ways to reduce the bounce rate \cite{SimilarWeb2019} by transferring the learning across different sites can help greatly level the playing fields. Section \ref{method_section} explains how to transfer context between \textbf{Site 1} and \textbf{Site 2} to condition the language model on \textbf{Site 2} for users which are at the very first interactions with the site itself. 

\begin{table}
  \caption{Most frequent queries by website section.}
  \label{tab:example_queries}
  \begin{tabular}{cccc}
    \toprule
    Rank & Ski & Women Sneakers & Man Apparel\\
    \midrule
    1  & ski trousers & saucony women & octopus sweater\\
    2  & man sweater & nike air force 1 & sweater\\
    3  & man ski trousers & air max 97 & octopus\\
    4  & ski gloves & nike air max & man sweater\\
    5  & ski jacket & running shoes & man overall\\
  \bottomrule
\end{tabular}
\end{table}

\textit{Coveo} is a SaaS search provider with more than 500 clients. As an industry case (see also our remarks in Section \ref{conclusion_section}), one of our clients has more than a dozen brands selling the same type of goods (i.e. shoes) and, at the same time, very little fine-grained historical interaction data are available. While planning the progressive roll-out of all brands, it became clear that deploying effective machine learning solutions at day one would have been unfeasible, highlighting the business value of applying scalable transfer learning strategies between the websites.

Our main research questions are therefore the following two:

\begin{enumerate}
    \item is there a fast-to-compute notion of “context” for e-commerce type-ahead that can be used to transform a “static” language model into a contextual one?
    \item can we map “context” from \textbf{Site 1} to \textbf{Site 2} in a totally unsupervised way to provide zero-shot personalization to new users on \textbf{Site 2}?
\end{enumerate}
 As we shall see, the answer is “yes” to both questions.

\section{Related works}
\label{related_section}
Suggest-as-you-type functionalities are very common in industry products and type-ahead has been a fairly well studied problem in academia in the IR and NLP communities~\cite{Cai:2016:SQA:3099948}. With the increasing penetration of neural architectures and the successes of “neural” language models (as opposed to more traditional, Markov-style discrete models), recent works have focused on leveraging recurrent neural networks to predict the most likely completion when shoppers are typing: \cite{inproceedingsparj2017} introduces a character-based language model for query completion; \cite{conf/sigir/WangZMDK18} deploys a noisy channel model for prediction as well, but their language model is unconditioned and most of the research is focused on inference speed. The availability of web search (as opposed to \textit{product} search) logs in the public domain has led to work on personalized suggestions with dense vectors mainly based on linguistic features. \cite{jaech-ostendorf-2018-personalized} learns personalized language models by training user embeddings as part of the end-to-end model: our user items are instead learned from catalog features only; they are cheaper to compute and allow an elegant extension of the model to the \textit{cross-shop} scenario as well. \cite{Shao_2018} uses vector similarity to re-rank suggestions, but it relies on the availability of big query logs and even public datasets (Google News) in the target language. On the commerce side, a recent work from \textit{eBay}~\cite{DBLP:journals/corr/abs-1905-01386} embeds previous user queries (through \textit{fastText}) and then re-ranks suggestions accordingly: while we also suggest a retrieve and re-rank strategy, our personalization layer does not require linguistic resources.  

A recurring theme in the recent literature in type-ahead is the prediction of “unseen queries” (e.g. \cite{inproceedingsparj2017}): by leveraging neural models, it is argued that type-ahead prediction can be effectively seen as a real-time “language generation” task. From a product perspective, we disagree that “unseen queries” are an important test case for commerce search, let alone the most important one: the final goal for suggesting queries to shoppers is not improving completion click-through rate \textit{per se}, but improving conversion down the line. Since type-ahead predictions are linked to conversion events through the search engine behavior, even a “mind-reading” type-ahead technology - one that always guesses the query intended by the user - is by itself no guarantee of higher conversions (as many “OR”-configured full-text engines will see their performance degrading by adding keywords). In the absence of business KPIs (i.e. conversion rate), we maintain that benchmarking models on “unseen queries” is not particularly significant when making strategic product decisions (see Section \ref{conclusion_section} for the role of \textit{offline} language generation in our roadmap). 

The literature on language models is extremely vast: even just focusing on neural language models (\cite{NIPS2000_1839}, \cite{Mikolov2010}, \cite{Xie2017NeuralTG}) - or just conditional neural language models (\cite{Keskar2019CTRLAC}, \cite{Mikolov2012ContextDR}) -, the literature is growing fast. In digital commerce, the concentration of innovation in a few mono-brand players may have obscured the importance of transfer learning for a vast portion of the market: recently, \cite{Authors2019} was the first research to focus on aligning product embeddings and publish proprietary data.

To our knowledge, the current research is the first to explore unsupervised personalization in NLP tasks across digital shops and to provide quantitative evidence and qualitative support that the proposed techniques can successfully address this important business scenario.

\section{Dataset}
\label{dataset_section}
The target dataset comes from a sample of 3 months of historical data for two digital shops, \textbf{Site 1} and \textbf{Site 2}, two mid-size websites (revenues between 10 and 100 million dollars a year) in the sport apparel space. Collected data includes:
\begin{itemize}
\item \textit{Consumer data}: data is collected through a \textit{web pixel} in compliance with existing legislation~\cite{Voigt:2017:EGD:3152676}; for the scope of \textit{this} work, we retrieve product impressions from anonymized shoppers; each impression event contains the \textit{SKU} of the product, that is the product unique identifier according to a shop catalog.
\item \textit{Image data}: catalogs from \textbf{Site 1} and \textbf{Site 2} contain one or more images for each product; images are relatively high-quality (typically 1200x1200), and they are copied from the public URL to our own infrastructure for further processing.
\item \textit{Type-ahead logs}: data is collected through a type-ahead API that records what anonymized users are typing (each keystroke), what the user sees in the drop-down menu and what, if anything, gets clicked.
\item \textit{Search logs}: data is collected through a search API that records what anonymized users are searching - either through type-ahead or spontaneously. 
\end{itemize}

\begin{table}
  \caption{Descriptive statistics for data collected for \textbf{Site 1} and \textbf{Site 2}.}
  \label{tab:descriptive}
  \begin{tabular}{cccc}
    \toprule
    Website & Products & Click events & Queries (type)\\
    \midrule
    Site 1  & 9002 & 944643 &  21488\\
    Site 2  & 12918 & 983468 &  30252\\
  \bottomrule
\end{tabular}
\end{table}

Descriptive statistics for the final dataset can be found in Table \ref{tab:descriptive}; Table \ref{tab:session_sample} shows an anonymized session comprising all types of events; interested practitioners can find additional details on data ingestion and data processing in \cite{Authors2019}. As a final remark, \textbf{Site 1} and \textbf{Site 2} have similarly high bounce rates (approximately 50\%) and low ratios of recurring users (<8\% of customers have 3 or more visits in 12 months), confirming the business importance of being able to produce session vectors \textit{with minimal information about the user} and \textit{as early as possible} in the shopping journey. Since \textbf{Site 1} and \textbf{Site 2} are independent shops, they are a perfect limit case to test the robustness of \textit{cross-shop} personalization without making any strong assumption of similarity in catalog structure and product taxonomy: it is therefore striking that, without any formal tie between the shops, almost 3\% of all sessions in \textbf{Site 1} for the test month have a companion session from the same users on \textbf{Site 2} (the same metric for brands in within a single group is significantly higher).

\begin{table}
  \caption{A sample session for anonymous user browsing snowboards on \textbf{Site 1}.}
  \label{tab:session_sample}
  \begin{tabular}{cccc}
    \toprule
    Timestamp & Session Id & Event Type & Data\\
    \midrule
    1575207599232  & 0002bf6d... & view & \textit{SKU}=0206395\\
    1575207647306  & 0002bf6d... & suggest & \textit{q}=d\\
    1575207647718  & 0002bf6d... & suggest & \textit{q}=dr\\
    1575207651617  & 0002bf6d... & search & \textit{q}=drake got\\
  \bottomrule
\end{tabular}
\end{table}

\section{Type-ahead models}
\label{method_section}
As a starting point, we assume that a "naive" unconditional model for type-ahead is given by the standard \textit{noisy channel model} (NCM)  \cite{conf/sigir/WangZMDK18}. Given a set of candidates queries $q_1$, $q_2$, \ldots $q_n$ and typed prefix $t$, we score candidate $q$ according to the Bayes formula:

\begin{equation}
  \tag{NCM}
  P(q | t) \propto P(q) * P(t | q)
  \label{eqn:ncm}
\end{equation}

A big advantage of \ref{eqn:ncm} for commerce search is that it lets us trade-off plausible completions with plausible typos \cite{brill-moore-2000-improved} (more than 10\% of search queries in the dataset contain typos): $P(q)$ is the “language model”, encoding the fact that most users search for “shoes” and not for “sweaters”,  $P(t|q)$ is the “error model”, encoding the fact that “zh” is most likely a misspelling for “sh” and not a genuine prefix. We are making the simplifying but realistic assumption that $P(t|q)$ is dependent on non-contextual factors (device, language, keyboard layout, etc.) that can be pre-calculated, and focus instead on estimating $P(q)$.

Queries in Table \ref{tab:example_queries} show that we should modify $P(q)$ above to $P(q|c)$: while it is theoretically possible to use an n-gram model encoding contextual elements at the start of the query, data sparsity will make the estimates almost relying entirely on smoothing for the vast majority of context/query pairs (a medium-size e-commerce site has more than 20k products and at least 50k pages). We turn instead to dense models, where embedding vectors can be easily plugged in to modify in real-time the unconditioned estimate of the basic model: but what vectors can represent the context well enough?

\subsection{Data preparation and session vectors}
\label{preparation_session}
As the vast majority of shoppers are not recurring, it is crucial to have a representation of unseen users that depends only on the current shopping session. Product images have been shown to perform well (sometimes significantly better than text descriptions) in content-based recommender systems \cite{10.1007/978-3-030-13709-0_40}; moreover, exploratory work in cross-shop recommendations confirmed that images could be a powerful feature to aid with transfer learning \cite{Authors2019}. It is important to note that relying on images alone for session representation allows personalization without any assumption on product meta-data or linguistic resources: while \textit{adding} information, if and when available, is a promising path for optimization, the goal of \textit{this} work is to prove that something so minimal like product images are both very scalable and effective (see also Section \ref{conclusion_section} for additional remarks).

We first extract features from product images as found in \textbf{Site 1} and \textbf{Site 2} catalogs; a deep convolutional neural network \cite{Simonyan15} (\textit{fc2} layer from the \textit{vgg16} model) extracts 4096-dimensional vectors from each product image, representing general visual features. PCA is then used to reduce the dimensionality of the vectors to 50 (the 50 components explain approximately 75\% of the original variance). 

Images can be pre-processed offline as they are available through static catalog files, so no specific deep learning hardware is necessary for data preparation; at session time, as the user is browsing different products, a simple call needs to be made to update a \textit{session cache}, that will retrieve the pre-computed 50-dimensional vector for the product and add it to the list. As explained below, we employ two strategies to encode user session starting from these vectors: first, we take an "average pooling" approach, so that a session with product $p_1$, $p_2$, \ldots $p_n$ is represented in the model as the average of the vectors for $p_1$, $p_2$, \ldots $p_n$ (averaging vectors to represent user interests is common in the recommendation literature, as for example suggested in \cite{10.1145/2959100.2959190}); second, when leveraging a full-fledged encoder-decoder architecture, we let the model ingest the vectors for $p_1$, $p_2$, \ldots $p_n$ separately and automatically learn how to condition language generation based on them.

\subsection{A similarity model}
\label{similarity_model_section}
The first improvement to our unconditional model enjoys the benefit of efficient training and easy run-time deployment, at the cost of some simplification on the modelling side. The intuition here is that we can encode the meaning of candidate queries through the vectors of the products that are most frequently associated with them: if the representation is solid, at run-time we re-rank the unconditioned suggestions by the distance in the vector space between the current session vector (representing user latent intent) and candidate queries' vector representation. Using surrounding events to build query representation is also at the heart of the \textit{Search2Vec} \cite{10.1145/2911451.2911538} model: while philosophically similar, applying the skip-gram model directly to queries is possible only with a huge amount of search sessions; our proposal leverages product interactions as building blocks instead, exploiting the fact that in typical shops most sessions (80\% to 90\%) are browsing sessions without search events.

In particular, at \textit{training} time, we:

\begin{enumerate}
    \item pre-process images and build a map from products $P$ to their vector representation, $P \mapsto V$;
    \item retrieve in the historical search data all the products $p_1$, $p_2$, \ldots $p_n$ clicked after each of the query candidates in $Q$, with relative frequency.
    \item build a map from query candidates to their vector representation, $Q \mapsto V$, by retrieving from $P \mapsto V$ vectors for $p_1$, $p_2$, \ldots $p_n$, and take their weighted average using the frequencies as weights.
\end{enumerate}

At \textit{run-time}, for shopper $S$, we:

\begin{enumerate}
    \item update the session vector $SV$ in the \textit{session cache} every time $S$ interacts with a product $p$;
    \item retrieve the current session vector $SV$ from the \textit{session cache};
    \item retrieve the top \textit{N} query candidates $q_1$, $q_2$, \ldots $q_n$ for query prefix \textit{t} using the unconditioned language model;
    \item retrieve for $q_1$, $q_2$, \ldots $q_n$ their vector representation from $Q \mapsto V$; 
    \item re-rank $q_1$, $q_2$, \ldots $q_n$ by calculating the cosine similarity between each query vector and $SV$.
\end{enumerate}

\subsection{An image-captioning model}
\label{enc_dec_model_section}
Our second improvement to the unconditioned language model has roots in sequence-to-sequence learning: evaluating probabilities of query candidates based on products can be elegantly modeled with an encoder-decoder architecture, as introduced in \cite{Sutskever2014SequenceTS} for machine translation and extended to image captioning in \cite{Vinyals2014ShowAT}. To encode the current session, we tried the two different strategies outlined in Section \ref{preparation_session}: an "average pooling" version, where the encoder has only one vector, i.e. the session vector resulting from averaging the product vectors in the session, and an "explicit" version, where the encoder takes as input \textit{the sequence} of all product vectors in the session to generate its output, i.e. the encoded session representation. 

The model is an encoder-decoder architecture, where the encoder part reads the session information and passes the encoded representation to the decoder, and the decoder is a character-based language model producing strings of text conditioned on the input session representation. The architecture is straightforward: the decoder is built with a single LSTM layer with 128 cells, followed by an output fully-connected layer with a softmax activation. The output dimensionality corresponds to the total number of unique characters in the training data, including the start-of-sequence and end-of-sequence tokens. We set a fixed sequence length for the decoder by taking the maximum length of all queries in the training dataset. During training, the latest cell states of the encoder are passed to the decoder as its initial cell states. At each timestep, we use teacher forcing strategy to pass the target character, offset by one position, as the next input character to the decoder \cite{Williams89alearning}. 

For training, we use Adam optimizer with an initial learning rate of 0.001 and a decay of 0.00001. Mini-batch of 128 samples per batch is used for training. Cross-entropy loss is used to backpropagate the error and update the weights for our model; training is performed over a maximum of 100 epochs, with early stop and $patience=20$.
Once trained, the model can be used to score a given pair of input context and input query. At run-time, we used the same retrieve-and-rerank strategy mentioned above: we first fetch candidates from the unconditioned model, and then re-rank the top queries using the probabilities of the conditional one. Scoring new queries is done by feeding the context into the encoder and then pass the encoded context, along with an input query, into the decoder. Then, at each timestep, the output of the decoder gives the probability of a target character, given a previous character of the query. Finally, we take the sum of all the output log-probabilities as the final score of the context-query pair. It is mentioned in the literature that such sequence scoring process might favor shorter sequences over longer ones. To overcome this problem, we apply a length-normalization method by dividing the score of every input query by its length to the power of a real number $r$ (empirically, $r=0.7$ is a value that works well \cite{Wu2016GooglesNM}). From our observations, this length-normalization technique is a simple yet efficient method that allows our model to correct length bias and deliver a better performance.

\section{EXPERIMENTS}
\label{experiment_section}
Type-ahead evaluations on web search datasets typically involve a temporal train/test split, with disjoint time periods to ensure a robust evaluation. In the commerce search case, though, the product context makes it harder to trust pure quantitative measures: completion probabilities at $t_{n}$ affect search queries frequency, which in turn affect completion probability at $t_{n+1}$, triggering a "rich get richer" dynamics that vastly overestimates the performance of a frequency-based baseline (the exploration/exploitation dynamics is itself an interesting challenge for live type-ahead systems). While online A/B testing is certainly preferable, sometimes it is not possible to tamper with production systems; for this reason, we chose to have two validation methodologies in place, to combine quantitative metrics in an offline setting and qualitative relevance judgments from humans.

\subsection{Within-shop scenario}
\subsubsection{Quantitative evaluation}
The experimental setting follows a standard train/test separation: all the models are trained with data from June to August, and tested on completely unseen events sampled from September. Average results for \textbf{Site 1} and \textbf{Site 2} over 10 runs, testing 7500 randomly sampled queries each run, are reported in Table \ref{tab:quantitative_results_first}. We use the \textit{mean reciprocal rank} (\textbf{MRR}) as a standard measure from the auto-completion literature; with \textbf{MRR@k} we indicate \textbf{MRR} as computed with the models returning at most $k$ candidate. In our evaluation, $k=5$ is picked to be consistent with the target production environments of \textbf{Site 1} and \textbf{Site 2}:

\begin{equation}
\mathrm{MRR}=\frac{1}{|Q|} \sum_{i=1}^{|Q|} \frac{1}{\operatorname{rank}_{i}}
\label{eq:mrr}
\end{equation}
where ${\operatorname{rank}_{i}}$ is the position of the first relevant result in the ${i}$-th query and ${Q}$ is the total number of queries. Results are reported for different \textit{seed} length, where \textit{seed} is the characters typed by a user into the search bar, i.e. query prefix, based on which our type-ahead models will be triggered to make suggestions: a seed length of $0$ corresponds to the scenario in which the user has just clicked on the search bar, but still has not typed anything; a seed length of $1$ is instead testing the suggestions of the given models after one typed character.

The models in the benchmark are the following:

\begin{itemize}
    \item \textbf{Popularity}: standard popularity ranking (e.g.  \textit{MPC} in \cite{Cai:2016:SQA:3099948}), where $P(q)$ is estimated from empirical frequencies in search logs;
    \item \textbf{Markov}: session is mapped to one out of a set of pre-defined buckets based on the activity type of the products in the sessions (e.g. \textit{tennis}, \textit{soccer}, etc.); type is extracted from the catalog through regular expressions with the help of product experts. The model is a bi-gram based model with Laplace smoothing;
    \item \textbf{Similarity}: cosine similarity re-ranking, as explained in Section \ref{similarity_model_section};
    \item \textbf{Enc-Dec "Avg"}: encoder-decoder architecture, using the "average" vector to represent the session, as explained in Section \ref{enc_dec_model_section};
    \item  \textbf{Enc-Dec "Full"}: encoder-decoder architecture, using the full list of product vectors to represent the session, as explained in Section \ref{enc_dec_model_section}.
\end{itemize}

\begin{table}
  \caption{MRR@5 at different seed length for models on \textbf{Site 1} and \textbf{Site 2}.}
  \label{tab:quantitative_results_first}
  \begin{tabular}{llcc}
    \toprule
    Shop & Model & Avg. (SD) Seed=0 & Avg. (SD) Seed=1\\
    \midrule
    \textbf{Site 1} & Popularity & 0.0134 (0.0006) & 0.0873 (0.001)\\
    & Markov & 0.011 (0.0004) & 0.071 (0.001)\\
    & Similarity & 0.029 (0.001) & 0.121 (0.001)\\
    & Enc-Dec "Avg" & 0.0376 (0.003) & \textbf{0.136 (0.005)}\\
    & Enc-Dec "Full" & \textbf{0.040 (0.002)} &  0.136 (0.006) \\
    \textbf{Site 2} & Popularity & 0.005 (0.0005) & 0.0982 (0.002)\\
    & Markov & 0.007 (0.0005) & 0.081 (0.002)\\
    & Similarity & 0.026 (0.002) & 0.109 (0.001)\\
    & Enc-Dec "Avg" & 0.0374 (0.003) & \textbf{0.147 (0.007)}\\
    & Enc-Dec "Full" & \textbf{0.0413 (0.003)} & 0.147 (0.008)\\
  \bottomrule
\end{tabular}
\end{table}

Baselines (other than being consolidated industry practices) respect the same data constraints that hold for the proposed models: they don't require linguistic resources and don't make strong assumptions on available product meta-data (see Section \ref{conclusion_section} for future developments).

Results are robust across sites, showing that dense personalized models significantly outperform popularity and discrete benchmarks: in the most common case of one-character seed, we observe 50\% increase in \textbf{MRR} between \textit{Popularity} and \textit{Enc-Dec}. Unsurprisingly, the less linguistic information is available (as in the case of predicting queries before any typing), the bigger the gap between the models is. The \textit{Similarity} model, despite its simplicity, is able to capture \textit{some} of the user intent, and it generally provides accuracy in between the baseline and the full deep learning model. It is also interesting to note that the difference between the "average" and the full-sequence models is negligible, perhaps due to the fact that most sessions are relatively short and long-range dependencies may not matter much. To give a sense of the generated predictions, Table \ref{tab:query_within_sample} collects sample of top queries from non-personalized vs personalized models.

\begin{table}
  \caption{Example of type-ahead suggestions (queries are translated) from \textbf{Site 2}, for a given prefix and a given session (represented with a significant product: running shoes, tennis racquets, etc). }
  \label{tab:query_within_sample}
  \begin{tabular}{lccc}
    \toprule
    Product & Seed & Popularity & Enc-Dec "Avg"\\
    \midrule
     \parbox[c]{1em}{\includegraphics[width=0.4in]{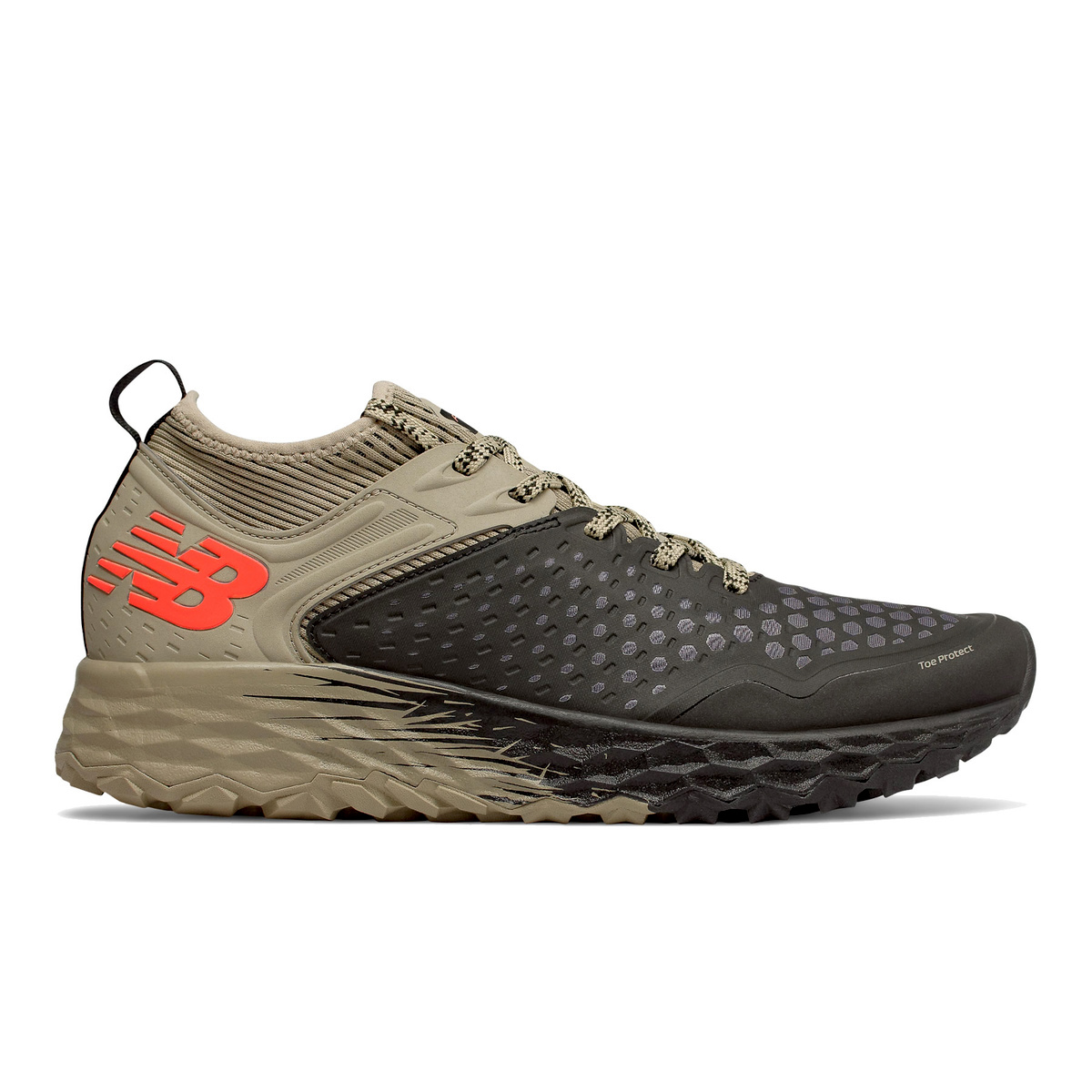}} & \textit{s} & trekking shoes & running shoes\\
    \parbox[c]{1em}{\includegraphics[width=0.4in]{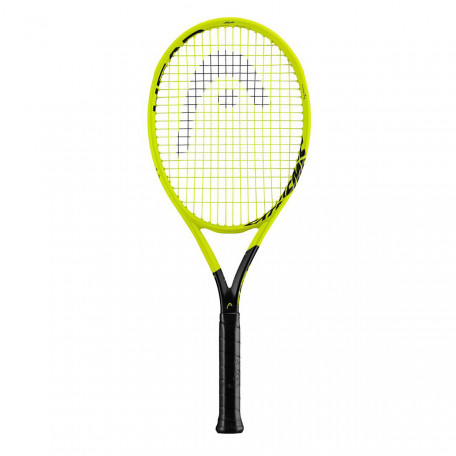}} & \textit{r} & reebok crossfit & tennis racquet\\
    \parbox[c]{1em}{\includegraphics[width=0.4in]{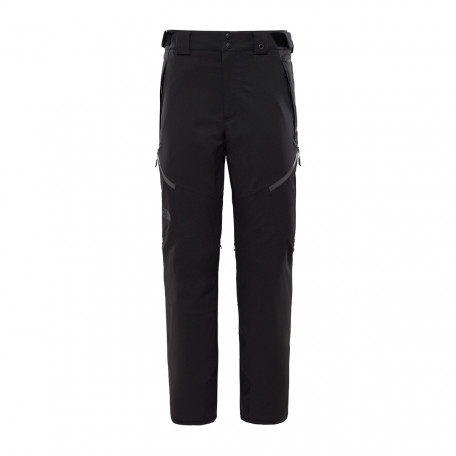}} & \textit{p} & men's polo & men's pants\\
    \parbox[c]{1em}{\includegraphics[width=0.4in]{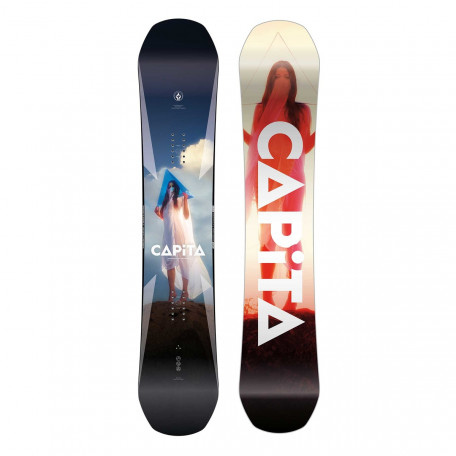}} & \textit{t} & t-shirt & snowboard\\
  \bottomrule
\end{tabular}
\end{table}

As noted in \cite{10.1145/1390334.1390364}, in the context of web search, not all types of queries benefit from personalization: for some search queries, most users are looking for the same thing; for others, different users want different results even if they use the same wording. To try and capture this effect in the context of product search, we leverage available dense representations instead of using a point-wise metric such as click entropy \cite{10.1145/1390334.1390364}: a large click entropy means many products are clicked in response to a given query, but it fails to capture the fact that, say, three running shoes are \textit{conceptually closer} in the "product space" than one t-shirt, one soccer ball, one pair of socks in response to the query "nike". In particular, we train \textit{prod2Vec} embeddings for SKUs in the dataset (\cite{Authors2019}) and calculate a dispersion value for each of the query \textit{q} in the candidate set, based on how tight is the cluster of products associated with it (sum of all distances from the centroid); we then assign all queries in the test set to one of two classes based on dispersion above/below the median value, and evaluate 5 runs of the \textit{Similarity} model for each class. The results greatly confirmed the web search prediction (\textbf{MRR@5}: 0.18 vs 0.034 for \textit{highly dispersed} vs \textit{tightly clustered queries}), confirming that personalized models are at their best dealing with ambiguous cases (i.e. cases in which linguistic information alone is not sufficient to determine product intent). These findings also open interesting lines of work for future refinements, in which a mixed model can be devised to treat differently ambiguous and non-ambiguous query candidates.

On a further note, recent papers have argued that deep learning models in NLP are both inefficient and unfair \cite{Strubell2019EnergyAP}, as the carbon footprint by GPUs is high and cloud costs make developing research papers prohibitively costly for data scientists outside a few tech companies. While we do not take any stance here over these arguments, we \textit{do} think that energy/cost/engineering considerations play a fundamental role in industry settings when calculating the right trade-off for the business: in this spirit, we find important to remark that our cosine similarity model can be trained and deployed successfully without devoted hardware (Appendix \ref{eng_appendix}).

\begin{figure}[h]
  \centering
  \includegraphics[width=\linewidth]{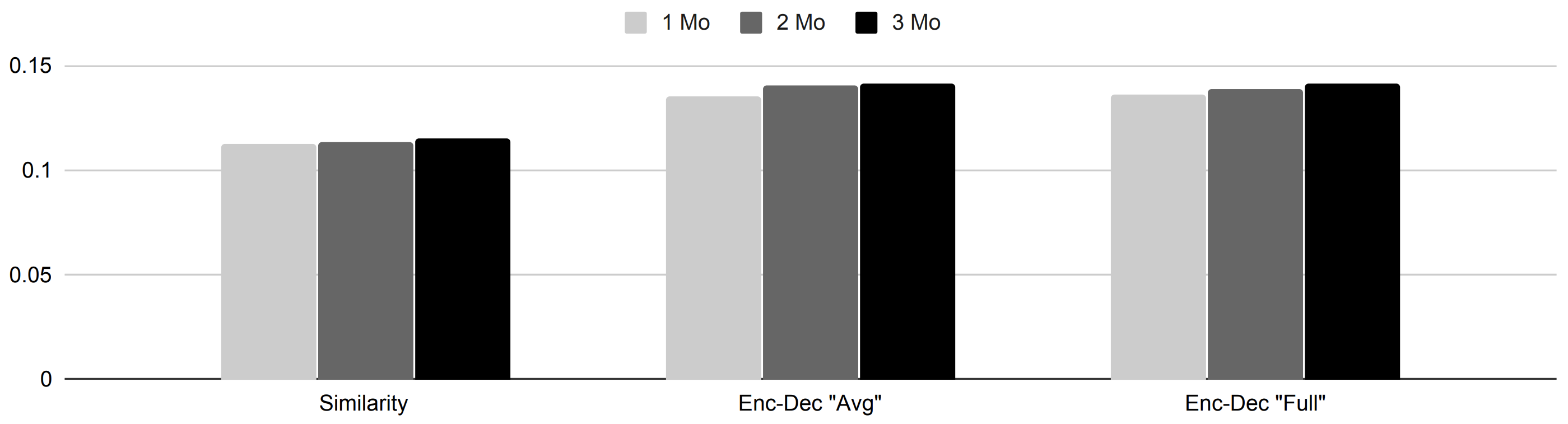}
  \caption{Average MMR@5 for 1/2/3 months of training data for the three dense models. }
  \label{training_pic}
  \Description{Type-ahead services need personalized ranking to be relevant.}
\end{figure}

Finally, we measured the robustness of the dense methods as the number of samples in the training set changes: the chart in Figure \ref{training_pic} show how the performances change as more data is included in the training set (from 1 to 3 months of data): not surprisingly, the encoder-decoder methods gain the most in accuracy as training size gets larger, but all dense models are better than baselines after only one month.

\subsubsection{Qualitative evaluation}
\label{qual_eva_section}
\begin{figure}[h]
  \centering
  \includegraphics[width=6.50cm]{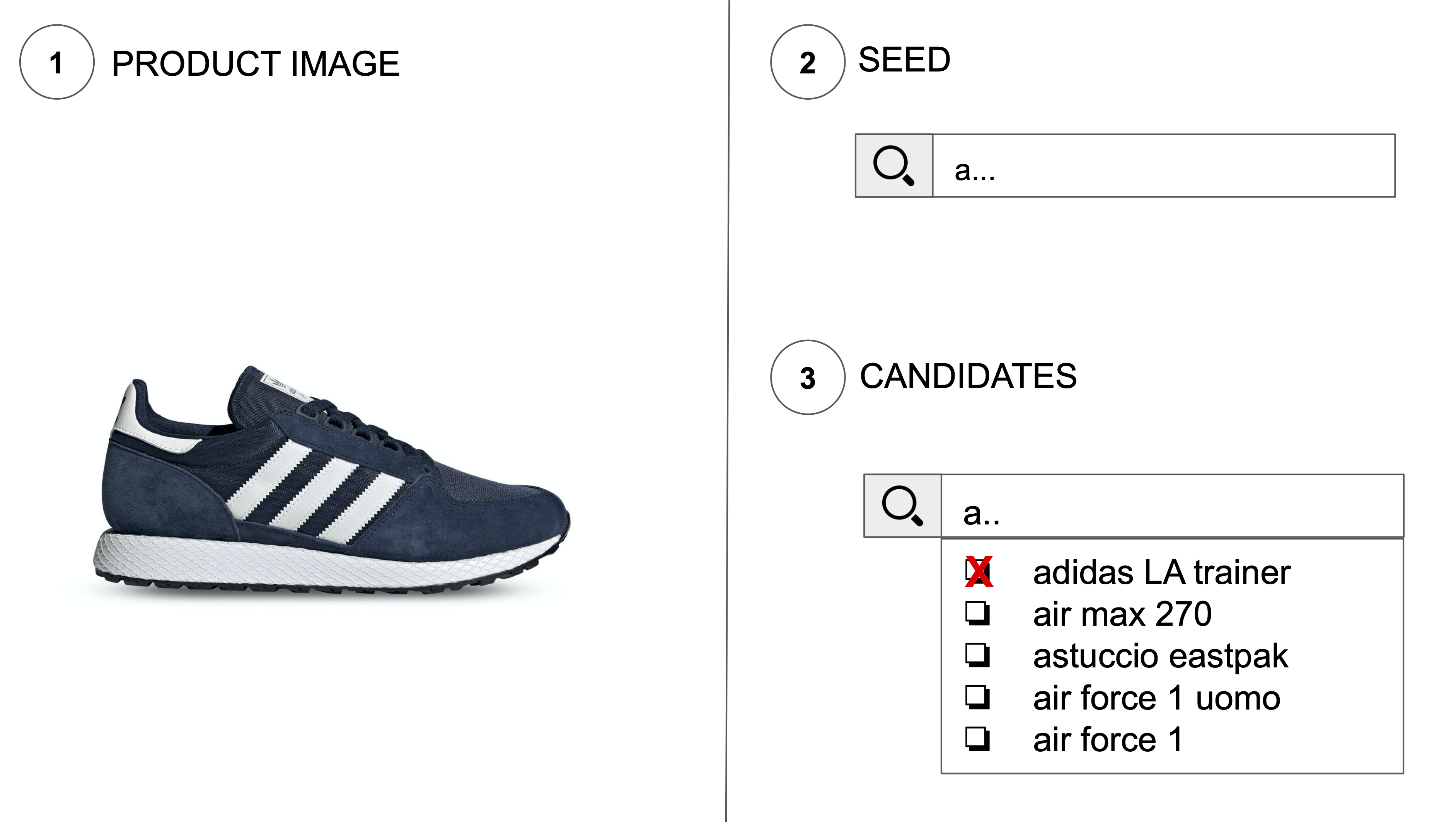}
  \caption{Example of the relevance task: subjects are presented with a product image, a seed and a list of query candidates, and they are asked to mark (red cross) the query that is deemed most appropriate for the product. }
  \label{qual_validation_pic}
  \Description{Example of the relevance task for qualitative evaluation.}
\end{figure}

To make sure the personalized models are actually capturing a meaningful difference in the "conceptual" space of human speakers, we set up a further validation procedure to avoid relying entirely on the quantitative measure. We recruited 15 native speakers who have no affiliation with \textit{Coveo}, \textbf{Site 1} or \textbf{Site 2} and whose age ranged between 22 and 45. We presented each of them with a series of \textit{stimuli} such as the one in Figure \ref{qual_validation_pic}.

The subjects were asked to pick the most relevant completion among 5 candidates, given a commerce product (represented by its image) and a seed; if no suggestion was deemed relevant, subjects were allowed to leave the form blank. The <\textit{product image}, \textit{seed}> pairs are taken from representative queries extracted from the training set for \textbf{Site 2}, for a total of 30 stimuli for each subject; candidate queries are chosen by first retrieving the top 25 candidates from the unconditioned model, and then sampling 5 times without replacement. 

By collecting relevance judgment from native speakers \textit{outside} the suggest-and-search loop, our prediction is two-fold: first, dense model should be confirmed to be relevant since conditional models are meant to capture important aspects of the semantic similarity between products in session and queries; second, the performance difference in \textbf{MRR} between the baseline and dense models should be higher, since the user study is meant to eliminate the popularity bias implicit in the search logs. After collecting the relevance judgments, we test \textit{Popularity} vs \textit{Cosine Similarity}, calculating the \textbf{MRR} over this new test set: \textbf{MRR@5} for \textit{Cosine Similarity} represents a 34.7\% increase over the baseline ($0.31$ vs $0.23$), which confirms both predictions; in particular, the purely quantitative evaluation for \textit{Site 2} showed only a 10.99\% increase over the baseline ($0.0982$ vs $0.109$). While the present user study was limited in both sample size and scope of semantic intuitions, we believe it provides supporting evidence to the main quantitative evaluation and it could potentially play even bigger roles in understanding specific strength and weaknesses of NLP personalization across different scenarios. For this reason, we look forward expanding this protocol to a larger user study for further iterations of this work.

\subsection{Cross-shop scenario}
To obtain a solution to the “zero-shot” personalization challenge in Figure \ref{pred_pic}, we need to map a session vector for \textbf{Site 1} to a session vector for \textbf{Site 2}: while product images obviously differ between categories \textit{within} shops and even more \textit{across} shops, the pre-processing pipeline is leveraging abstract features as extracted from a general purposes convolutional neural network that is expected to generalize to many types of objects, photographed under very different conditions. Since product images live all in this general and noise-tolerant "image space", our hypothesis is that zero-shot personalization can be achieved by combining the vectorization component of a model trained on \textbf{Site 1} with the language model trained on \textbf{Site 2}. 

To test the hypothesis, we run quantitative benchmarks on the cross-shop portion of our dataset. Results for the cross-shop personalization scenario are reported in Table \ref{tab:quantitative_results_second}, comparing a popularity model where shoppers landing on a new website receive a non-personalized prediction with two personalized models (\textit{Similarity} and \textit{Enc-Dec "Avg"}) injecting a session vector computed from the \textit{previous} site. 

The test set is composed by sessions with $n>0$ products viewed on \textbf{Site 1} and then a matching session, the same day, on \textbf{Site 2}  with $n>0$ products viewed and one search query issued (we require products to be present on the target site as well to allow for a within-site comparison). Quantitative evaluation for the test period is done on all the 2137 sessions matching the above conditions.   
\begin{table}
  \caption{MRR@5 ($seed=1$) for shoppers browsing on two shops.}
  \label{tab:quantitative_results_second}
  \begin{tabular}{lcc}
    \toprule
    Model & MRR@5 {1}->{2} & MRR@5 {2}->{1}\\
    \midrule
    Popularity & 0.048 & 0.086\\
    Cross-shop Similarity & 0.060 & 0.093\\
    Cross-shop Enc-Dec & \textbf{0.083} & \textbf{0.103}\\
    Within-shop Similarity & 0.096 & 0.127\\
  \bottomrule
\end{tabular}
\end{table}

By transferring the session vector between sites, we obtain a significant increase for \textbf{MRR@5} over an industry benchmark that treats the incoming shopper as new. To show how hard the task is, we include as a "conceptual" upper bound a \textit{Site Similarity} model, which is a model trained on the target site and using same site context on the test set: it is remarkable that the intent is transferred so well between sites that the best dense model with transfer learning is very close in performance to a same-site model. While the personalization accuracy \textit{across sites} is indeed lower than \textit{within shop}, the main concern of \textit{this} work is proving that transferred intent is significantly better than assuming every shopper is new on the target site; for online shops making close to 100 million dollars in revenue/year or more, capturing the interest of even a small percentage of the portion of non-recurring users may provide significant business benefits.

We also tested the hypothesis that cross-shop sessions closer in time would further improve accuracy for the transfer learning model: if we require the shopper to complete the sessions on the two shops in less than one hour, the gap between unconditioned vs conditional model widens ($0.48$ for \textit{Popularity} vs $0.72$ for \textit{Similarity} from \textbf{Shop 1} to \textbf{Shop 2}), but test set size becomes too small to make definite conclusions on this specific hypothesis - we look forward repeating the quantitative evaluations when more cross-shop data becomes available.

It is worth highlighting that by using the shared image space, we do not need to change our language model or re-train it: models for \textbf{Site 1} and \textbf{Site 2} can be trained and deployed independently; the only change required to unlock zero-shot personalization in a production environment is that, in the absence of history on the target shop, the language model gets injected with a “transferred” session vector. 

\section{Conclusion and Future Work}
\label{conclusion_section}
\textit{This} work successfully shows that it is possible to leverage easy-to-compute image features to obtain fast and scalable dense representations for commerce products. Starting from an unconditioned noisy channel, we show how to inject personalization with session vectors. The effectiveness of the proposed strategies has been benchmarked with quantitative measures supplemented by a qualitative study, designed to obtain a more nuanced evaluation of the semantic quality of the personalized rankings. Prompted by real industry use cases, we generalized the models to the "cross-shop" scenario, in which shoppers move across similar websites and personalization need to be provided by transferring learned intent from one shop to another. To the best of our knowledge, \textit{this} is the first work to explicitly address both \textit{within} shop and \textit{across} shops scenarios in auto-completion personalization.

It is important to remark that the proposed strategy enjoys several advantages over alternative proposals:

\begin{itemize}
    \item personalization features - i.e. image vectors - are easy to compute, do not need frequent updates (as product images change rarely) and do not require data from the user except within-session interactions: this makes the model very effective for mid-size shops where the majority of users is not recurring and personalization needs to be achieved with as little data as possible;
    \item our methods do not require catalogs to be particularly accurate in their description, or comparable in their structure: as all digital shops have images for products, the methods presented enjoy widespread, immediate applicability in a retail industry where data quality is not always ideal;
    \item our methods do not require advance tracking in place to start, making it ideal in integration scenarios in which historical data tracking is incomplete;
    \item our methods are privacy friendly, as they do not require storing browsing histories for all shoppers: the data in the session cache can expire automatically after most sessions as long as some are kept to continue with model training. At a time of increasing concern for data regulations \cite{Voigt:2017:EGD:3152676}, being able to provide personalized experiences without storing long-term behavioral data on specific users may be an important feature for players in the industry;
    \item our methods do not require language resources, which is often a problem for mid-size, multi-lingual commerce sites: on the one hand, linguistic resources (say, word vectors \cite{pennington-etal-2014-glove}) are often not available for all languages; on the other, most shops do not have enough textual data to train effective vectors themselves;
    \item our methods provide a principled way to transfer learning across sites, which is one of the biggest challenges for e-commerce service providers and a massive opportunity for multi-brand retailers, which often need to transfer learning from data-rich stores to “zero-data” stores;
    \item finally, the methods allow for a step-wise implementation process, which allows for incremental change to standard industry type-ahead APIs (see Appendix \ref{eng_appendix} for engineering details).
\end{itemize}

While our results are very promising, our research opened up many possibilities for further improvement, both from a product and research standpoint. The deployment of full-fledged in-session personalization on all NLP touchpoints is one of the most important items in \textit{Coveo}'s product roadmap: in particular, our multi-brand groups will be involved in all the product choices (data ingestion, real-time processing, etc.) needed to deploy personalization at scale. We are also investing heavily in improving engineering and model serving, making sure the uplift in performances and elegance gained with dense models is sustainable at scale in a fast growing organization.

Our research roadmap on type-ahead mainly focuses on two themes. For reasons mentioned in Section \ref{related_section}, we are skeptical of unconstrained real-time natural language generation, potentially resulting in unseen queries with low business value (i.e. conversion rate) and unnecessarily complex serving infrastructure. However, we do believe a big goal of type-ahead is helping \textit{discovering} portions of the product space the users may not even know exist. To that extent, natural language generation as performed \textit{offline} at indexing time is a powerful technique to increase the number of query candidates in a controllable way. A second point is obviously the conditioning process, as product images are easy to compute, but their representation potential may be augmented by other unsupervised representations, ranging from behavioral-based product embeddings~\cite{Authors2019}, to page embeddings (i.e. not necessarily product pages), to longer-term user information (when available): \textit{this} work proved that image-based deep architectures significantly improve performances of type-ahead services, however it is still an open question which combination of dense representations and deep architectures exactly perform the best. We leave this point to future research.

Finally, it is our deep conviction that transfer learning is a core component of any commerce solution with the ambition of serving hundreds-to-thousands of customers in a global market: for this reason, the understanding of "cross-shop" behavior is an active area of research in our lab, from recommendation~\cite{Authors2019} to several NLP tasks.

\begin{acks}
Thanks to Luca Bigon for support on data ingestion and engineering tooling needed for the project; thanks to three anonymous reviewers, whose feedback greatly improved the paper. Finally, thanks to our clients, who are instrumental in the success of the company and have been very receptive to the possibilities opened by A.I. in retail.
\end{acks}

\bibliographystyle{ACM-Reference-Format}
\bibliography{short_refs}

\appendix

\section{Architectural notes}
\label{eng_appendix}
As specified when formulating \ref{eqn:ncm}, we assume as a baseline implementation a "naive" unconditioned language model based on popularity and deploy a retrieve-and-re-rank strategy, first retrieving suitable candidates and then scoring them with the \textit{conditional} language model we learned from the data. Assuming we can generate the list of candidate queries in advance (see also note on query generation in Section \ref{conclusion_section}), our starting implementation is a trie-based architecture leveraging the fact that both the language model and the error model can be estimated offline when building the index. An example of this data structure for indexing query candidate \textit{shoes} is depicted in Figure \ref{basic_arch_pic}:

\begin{figure}[h]
  \centering
  \includegraphics[width=\linewidth]{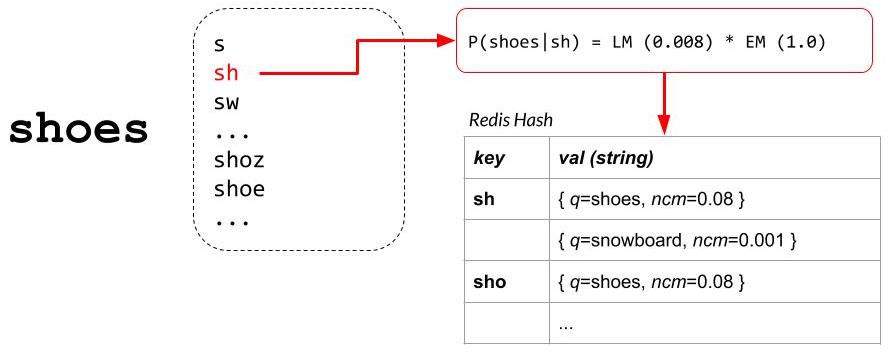}
  \caption{Indexing \textit{shoes} in a trie built on top of an hash-map, pre-calculating at indexing time the unconditioned probabilities for all the possible completions of a given prefix.}
  \label{basic_arch_pic}
  \Description{Basic trie-based storage.}
\end{figure}

At run-time, we retrieve from the trie the top \textit{N} completions for the given prefix, since they are stored in descending probability order (this data structure allows us to trade off space for time, resulting in constant query time thanks to hash computational complexity\footnote{\url{https://redis.io/commands/hget}}).

\begin{figure}[h]
  \centering
  \includegraphics[width=\linewidth]{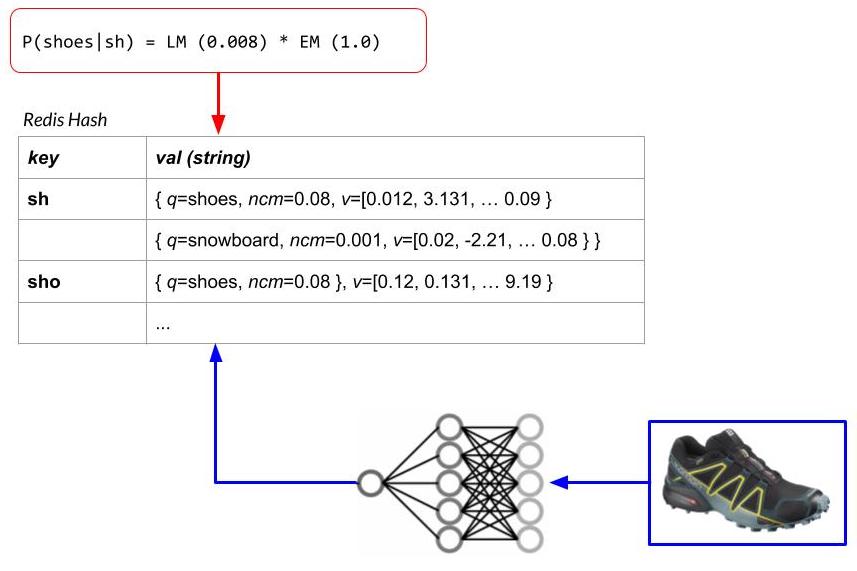}
  \caption{When dense vectors are introduced, indexing \textit{shoes} in a trie structure requires merging historical search data with the results of the image processing pipeline.}
  \label{similar_arch_pic}
  \Description{Similarity trie-based storage.}
\end{figure}

We can progressively enhance this setup by introducing in steps our two personalization models. First, we can augment the indexing process \textit{without} deploying expensive hardware with another offline process - image vector calculations - to prepare for the similarity-based personalization; we first retrieve for \textit{shoes} images of products clicked by users issuing the query, run them through the CNN and store the resulting vector in the same JSON representing the property of the completions in the Redis hash (Figure \ref{similar_arch_pic}). At run-time, we retrieve the top $N$ * $k$ completions and re-rank them based on the distance between their vectors representation and the current session vector, as retrieved by a session cache: to maintain the \textit{session cache}, the easiest solution would be to adopt a lightweight memory store updated at each product view. 

\begin{figure}[h]
  \centering
  \includegraphics[width=\linewidth]{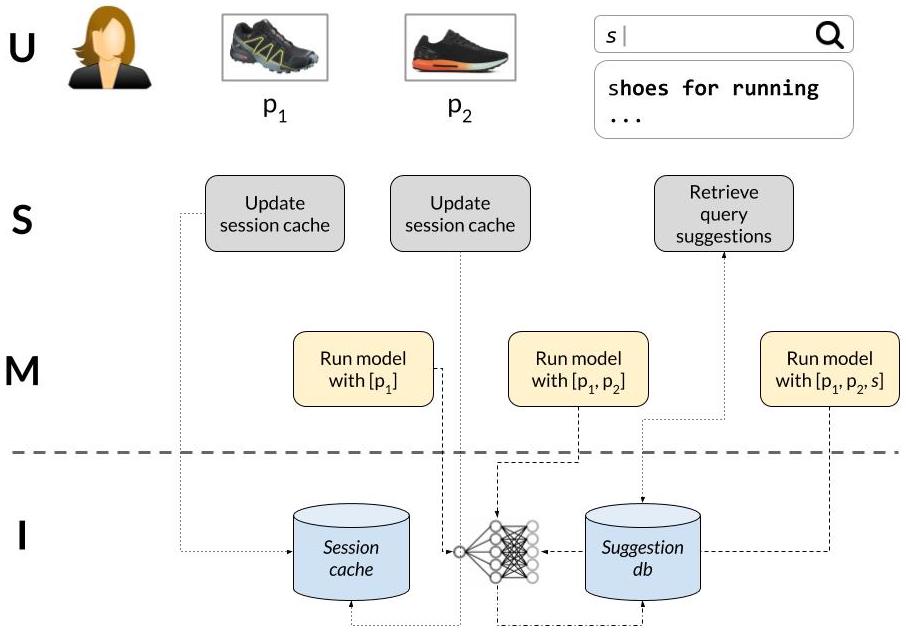}
  \caption{Shopper, website and model timeline together in an hybrid system that is achieving personalization in "a best effort" mode: as the shopper moves on the target domain, neural networks behind the scene adjust probabilities for candidate queries, decoupling run-time performance of the API from the personalization component.}
  \label{encoder_arch_pic}
  \Description{Deep learning trie-based storage.}
\end{figure}

We maintain that the simplicity of this extension is an argument in favor of the similarity model, but we would still like to find ways to extend the architecture to accommodate the encoder-decoder model with minimal disruption. In particular, we focus on a solution that allows us to decouple deep learning models for personalization from query time constraints - a solution that degrades gracefully in case of any problem to the deep learning serving layer. Our proposal for an "hybrid system" of this kind is depicted in Figure \ref{encoder_arch_pic}.

Figure \ref{encoder_arch_pic} shows three timelines - \textit{User}, \textit{Shop}, \textit{Model} - and some basic \textit{Infrastructure} resources at the bottom. Going from left-to-right, we follow the user journey through the website:

\begin{itemize}
    \item at each product view, the shop \textit{S} takes care of synchronously updating the cache;
     \item each cache update triggers an asynchronous model run (\textit{M} timeline) to estimate the conditional probabilities for the top $U$ queries (where $U$ is a much greater number than what is usually displayed to the user on the website), which are stored in a temporary hash in the query completion database;
     \item when the shopper finally types "s" in the search bar and the website needs to retrieve suggestion, it retrieves the top $N$ * $k$ completions, but re-rank them with the conditional probabilities for the queries as stored at the last model update. 
\end{itemize}

In other words, the model layer acts in a sort of "best effort" mode, trying to update as fast as possible query candidates probabilities to provide more accurate suggestions, knowing well that not \textit{every} candidate will be ranked correctly, but knowing that \textit{most} candidates will be without corrupting in any way the user experience on the website. An additional benefit of this hybrid approach is that the deep learning infrastructure can be introduced gently first and with minimal investment, and optimize later (both on hardware and software side) when ROI is more easily measured.

\end{document}